\RequirePackage{fix-cm}
\let\accentvec\vec
\documentclass[twocolumn]{svjour3}

\let\vec\accentvec

\usepackage{graphicx}
\usepackage[latin9]{inputenc}
\usepackage{float}
\usepackage{amsmath}
\usepackage{amssymb}
\usepackage{natbib}
\usepackage[colorlinks=true,allcolors=blue]{hyperref}
\usepackage{txfonts}

\usepackage[switch]{lineno}

\journalname{}

\begin{document}

\title{Miniaturized lab system for future cold atom experiments in microgravity
}

\author{Sascha Kulas         \and
        Christian Vogt          \and
        Andreas Resch         \and
        Jonas Hartwig       \and
        Sven Ganske        \and
        Jonas Matthias      \and
        Dennis Schlippert    \and
        Thijs Wendrich      \and
        Wolfgang Ertmer      \and
        Ernst Maria Rasel    \and
        Marcin Damjanic      \and
        Peter Weßels         \and
				Anja Kohfeldt        \and
				Erdenetsetseg Luvsandamdin \and
        Max Schiemangk      \and
        Christoph Grzeschik \and
        Markus Krutzik       \and
        Andreas Wicht      \and
				Achim Peters       \and
				Sven Herrmann      \and
        Claus Lämmerzahl
}

\institute{S. Kulas         \and
        C. Vogt          \and
        A. Resch         \and
        S. Herrmann \and
        C. Lämmerzahl
             \at
              ZARM, Universität Bremen, Am Fallturm, D-28359 Bremen, Germany \\
              \email{sascha.kulas@zarm.uni-bremen.de}  
           \and
           J. Hartwig \and
           S. Ganske  \and
           J. Matthias \and
           D. Schlippert \and
           T. Wendrich \and
           W. Ertmer \and
           E. M. Rasel
           \at
              Institut für Quantenoptik and Centre for Quantum Engineering and Space-Time Research (QUEST),
Leibniz Universität Hannover, Welfengarten 1, D-30167 Hannover, Germany \\
    \and
           M. Schiemangk  \and
           C. Grzeschik \and
           M. Krutzik \and
           A. Wicht \and
           A. Peters
           \at
              Institut für Physik,
Humboldt-Universität zu Berlin, Newtonstraße 15, D-12489 Berlin, Germany\\
    \and
		A. Kohfeldt  \and
		E. Luvsandamdin \and
    M. Schiemangk \and
     A. Wicht \and
     A. Peters
     \at
     Ferdinand-Braun-Institut, Leibniz-Institut für Höchst\-frequenztechnik,
Gustav-Kirchhoff-Straße 4, D-12489 Berlin, Germany\\
   \and
   M. Damjanic      \and
   P. Weßels
    \at
     Laser Zentrum Hannover e.V., Hollerithallee 8, D-30419 Hannover, Germany\\
}

\date{Received: date / Accepted: date}

\maketitle

\begin{abstract}
We present the technical realization of a compact system for performing experiments with cold $^{87}{\text{Rb}}$ and $^{39}{\text{K}}$ atoms in microgravity in the future. The whole system fits into a capsule to be used in the drop tower Bremen. One of the advantages of a microgravity environment is long time evolution of atomic clouds which yields higher sensitivities in atom interferometer measurements.
We give a full description of the system containing an experimental chamber with ultra-high vacuum conditions, miniaturized laser systems, a high-power thulium-doped fiber laser, the electronics and the power management. In a two-stage magneto-optical trap atoms should be cooled to the low $\muup$K regime. The thulium-doped fiber laser will create an optical dipole trap which will allow further cooling to sub-$\muup$K temperatures. The presented system fulfills the demanding requirements on size and power management for cold atom experiments on a microgravity platform, especially with respect to the use of an optical dipole trap. A first test in microgravity, including the creation of  a cold Rb ensemble, shows the functionality of the system.
\end{abstract}

\section{Introduction}
\label{intro}
In 1924 de Broglie realized that propagating particles can be described as waves and formed the theoretical basis for matter-wave interferometers (de Broglie \citeyear{Broglie1924}). The interference of matter-waves was later demonstrated in various proof-of-principle experiments (Davisson et al. \citeyear{Davisson1927}; Jöns\-son \citeyear{Joensson1961}). Then in 1991, Riehle presented the first gyroscope based on light pulse matter-wave interferometry at the PTB in Braunschweig, Germany (Riehle et al. \citeyear{Riehle1991}) and shortly afterward Kasevich and Chu built the first gravimeter with cold atoms (Kasevich et al. \citeyear{Kasevich1992}). Since then the interest in interferometry with matter-waves has been steadily rising and several fundamental measurements with very high sensitivity have been performed, such as measurements of the fine structure constant (Bouchendira et al. \citeyear{Bouchendira2013}; Wicht et al. \citeyear{Wicht2002}) and of the gravitational constant (Lamporesi et al. \citeyear{Lamporesi2008}) or tests of the weak equivalence principle (Zhou et al. \citeyear{Zhou2015}; Schlippert et al. \citeyear{Schlippert2014}).\newline
Matter-wave interferometers are based on coherent manipulation of matter with light pulses. The manipulation is done by three subsequent two-photon Raman transitions where the internal hyperfine state is changed and two photon momenta are transferred. In a Mach-Zehnder setup the first light pulse creates a coherent superposition of two atomic wave\-packets propagating along different paths. After a time $T$ a second laser light pulse redirects the wave\-packets and after the time $2T$ a third light pulse causes them to interfere. Detection of the interferometer outputs allows to infer the leading order phase shift. In a three-pulse matter-wave accelerometer this phase shift is  
\begin{linenomath}
\begin{equation}
\label{phi}
\phi=kaT^2,
\end{equation}
\end{linenomath}
where $k$ is the laser wavevector and $a$ the acceleration of the atoms. To enhance the sensitivity by using long pulse separation times $T$ typically atomic fountains are used (Clairon et al. \citeyear{Clairon1991}; Kasevich et al. \citeyear{Kasevich1992}). Even longer pulse separation times can be achieved in microgravity environments.

Moving experiments to microgravity is pursued in various activities worldwide. The QUANTUS collaboration has de\-monstrated interferometry using Bose-Einstein condensates (BECs) in drop-tower experiments (Müntinga et al. \citeyear{Muentinga2013}), currently working towards a first space-based BEC experiment on board a sounding rocket (Seidel et al. \citeyear{Seidel2013}). The French ICE project has reported on matter-wave inter\-ferometry on board the zero-g Airbus (Geiger et al. \citeyear{Geiger2011}), and the satellite mission STE-QUEST was studied as a candidate for an M-class mission in the European Space Agen\-cy's Cosmic Vision program (Altschul et al. \citeyear{Altschul2015}; Aguilera et al. \citeyear{Aguilera2014}). Finally, NASA plans the installation of a Cold Atom Lab on board the International Space Station for 2017 (http://coldatomlab.jpl.nasa.gov).

This work now aims to build a differential matter-wave accelerometer to be operated simultaneously with two atomic species. Here we report on the first required step herefore, which is the realization of a suitable source, a mixture of cold $^{87}{\text{Rb}}$ and $^{39}{\text{K}}$ atoms. Concerning cooling and interferometry aspects, $^{87}{\text{Rb}}$ is a well-known alkali metal. Its D2-line is in the visible infrared range (780 nm) and cooling and manipulation can be done using semiconductor laser diodes. $^{39}{\text{K}}$ has a similar wavelength of the D2-line (767 nm) which allows to use the same optics and easily combine light paths from both laser systems. Mixtures of cold $^{87}{\text{Rb}}$ and $^{39}{\text{K}}$ have been well studied previously (Roati et al. \citeyear{Roati2007}), and both species are also used for interferometry in laboratory experiments (Lamporesi et al. \citeyear{Lamporesi2008}; Fattori et al. \citeyear{Fattori2008}).

A major motivation for differential interferometry with K and Rb isotopes is to perform a comparison of their free fall, and thus a test of the weak equivalence principle underlying the theory of general relativity. For analysis of this experiment the Eötvös factor $\eta$ is typically used, which is essentially the normalized difference of free-fall acceleration $a_{\text{Rb/K}} = g_{\text{Rb/K}}$ of the two test masses. Using the above equation (\ref{phi}) we can express this as
\begin{linenomath}
\begin{equation}
\eta=2\,\frac{g_{\text{Rb}}-g_{\text{K}}}{g_{\text{Rb}}+g_{\text{K}}}=\frac{\delta \phi}{gS},
\end{equation}
\end{linenomath}
with $g=(g_{\text{Rb}}+g_{\text{K}})/2$, the differential interferometer phase $\delta \phi$ and equal interferometer scaling factor $S=k_{\text{Rb}}T^{2}_{\text{Rb}}=k_{\text{K}}T^{2}_{\text{K}}$. In ground-based experiments such tests have already been achieved with a recent experiment comparing the free fall of Rb and K atoms. This experiment obtained $\eta = (0.3 \pm 5.4)\times 10^{-7}$ with $T$ up to 20\,ms and averaging over approxi\-mately 4000\,s (Schlippert et al. \citeyear{Schlippert2014}). For the design of our microgravity experiment we take this experiment as a guideline to explore the potential gain provided by the extended pulse separation time $T$ in microgravity. If we assume a pulse separation time of $T=0.5$\,s we obtain $S=4\times 10^{7} {\text{rad}}/g$. Using ellipse fitting for suppression of common-mode phase fluctuations due to vibrations we aim to detect the differential phase at $\delta \phi< 100$\,mrad resolution which should allow for a
single-shot sensitivity of 
\begin{linenomath}
\begin{equation}
\eta_{\text{est}} < 2.5\times 10^{-9}.
\end{equation}
\end{linenomath}
A crucial aspect in our experimental setup is the plan to use an optical dipole trap in contrast to a macroscopic atom chip. Atom chips have so far been the primary choice in microgravity experiments with ultra-cold degenerate quantum gas\-es since they provide the obvious benefit of low energy consumption due to the small distance between chip and atoms. High magnetic-field gradients can be generated for efficient cooling processes. Potential disadvantages of an atom chip besides limited optical access are diffractions of the interferometer beam at the edge of the chip and distortions of the wavefronts. Other perturbations affecting precision measurements may arise from surface potentials and gravitational gradients. These can be avoided if an optical dipole trap is used (Zaiser et al. \citeyear{Zaiser2011}). Also, the preparation of ultra-cold mixtures of atoms will require the use of Feshbach resonances (Roati et al. \citeyear{Roati2007}), not compatible with a purely magnetic chip trap. Thus also atom chip experiments aiming for ultra-cold mixtures foresee combination with an optical trap at some point. An all-optical technique however is more demanding with respect to power management and size, which is an important criterion for microgravity experiments.

We use the drop tower Bremen as a microgravity platform. The whole system thus has to fit into a capsule with a payload volume of 1.7\,m in height and 0.7\,m in diameter. Drop tower operation is restricted to capsules of mass $<500\,\text{kg}$. With supporting structures of 279\,kg including platforms, stringers, hull and interface unit this leaves 221\,kg to the payload. At a height of 110\,m the drop tower provides a free-fall time of 4.7 seconds. Compared to other microgravity platforms such as parabolic flights in an airplane (Geiger et al. \citeyear{Geiger2011}) or sounding rockets, the drop tower Bremen offers easy access and excellent zero-g-quality with residual accelerations at the low $10^{-6}g$ level after evacuation of the drop tube to a pressure below 10\,Pa. Deceleration of the drop capsule is achieved within 0.2\,s within a container filled with styrofoam pellets. The experiment thus has to be designed such that it withstands a resulting peak deceleration of up to $40\,g$.

This paper is organized as follows: Section 1.1 gives a general overview of the system. Section 2 and 3 describe the basic outline of the system, i.e. the vacuum system and the laser setup. Section 4 gives the first experimental characteristics concerning the performance of the magneto-optical traps. Section 5 describes the implementation of the high-power laser to create the optical dipole trap and section 6 presents the first test in microgravity. Finally, we end with a conclusion and an outlook on the next steps.

\subsection{Structure of the system}
\label{sec:1}
\begin{figure}
\resizebox{0.48\textwidth}{!}{%
  \includegraphics{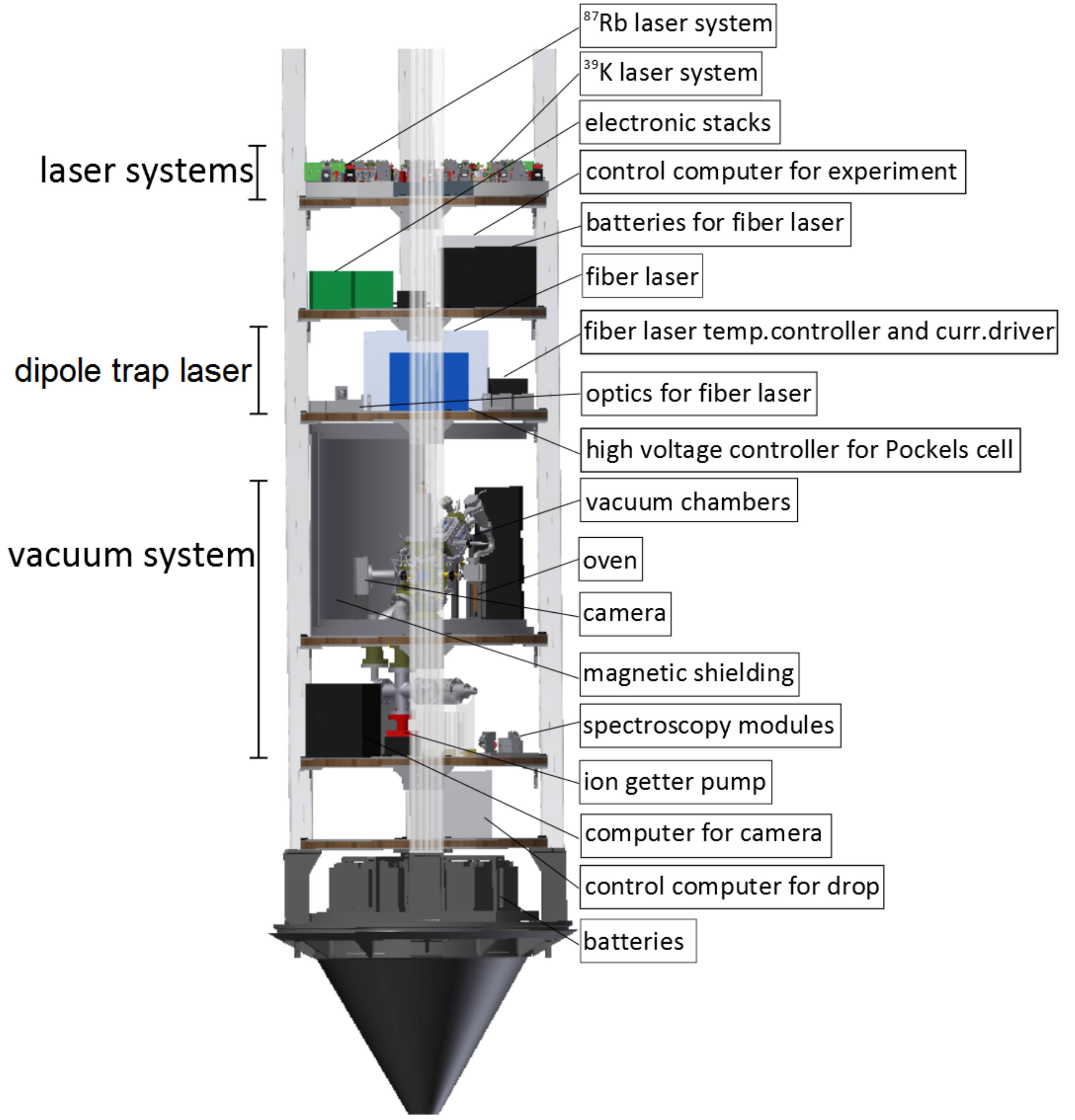}
}
\caption{Setup of the drop capsule. The magnetic shielding around the vacuum chambers is not shown completely for reasons of clarity.}
\label{Setup}
\end{figure}

Figure \ref{Setup} shows the whole system. At the heart are two vacuum chambers (located at the center of mass of the drop capsule) which are connected by a differential pumping tube. The smaller source chamber contains a ${\text{2D}}_{+}$-MOT (see Fig. \ref{vacuum chamber}) which is a combination of a 2D-MOT (two dimensional magneto-optical trap) and an additional pusher and retarder beam for better velocity selection (Dieckmann et al. \citeyear{Dieckmann1998}).
The main chamber is the experimental chamber which contains a 3D-MOT. The magneto-optical traps need so called cooling and repumping light to run the cooling cycle. These differ in frequency and intensity and are supplied by the two laser systems at the top of the capsule (see Fig. \ref{Setup}). The light is provided via optical fibers and telescopes attached to each viewport. The chambers are enclosed within a two-layer magnetic shielding. To create the optical dipole trap, an additional high-power continuous wave (cw) laser beam, far off red detuned at $2\,\muup$m wavelength, is focused into the center of the experimental chamber. This beam is generated by a thulium-doped fiber laser located on the platform above with a free-beam guide to the vacuum chamber. The electronic control stacks are housed above and a battery platform for autonomous power supply is located at the bottom of our system.
\section{Vacuum chambers}
\label{sec:2}
\begin{figure}
\resizebox{0.48\textwidth}{!}{%
  \includegraphics{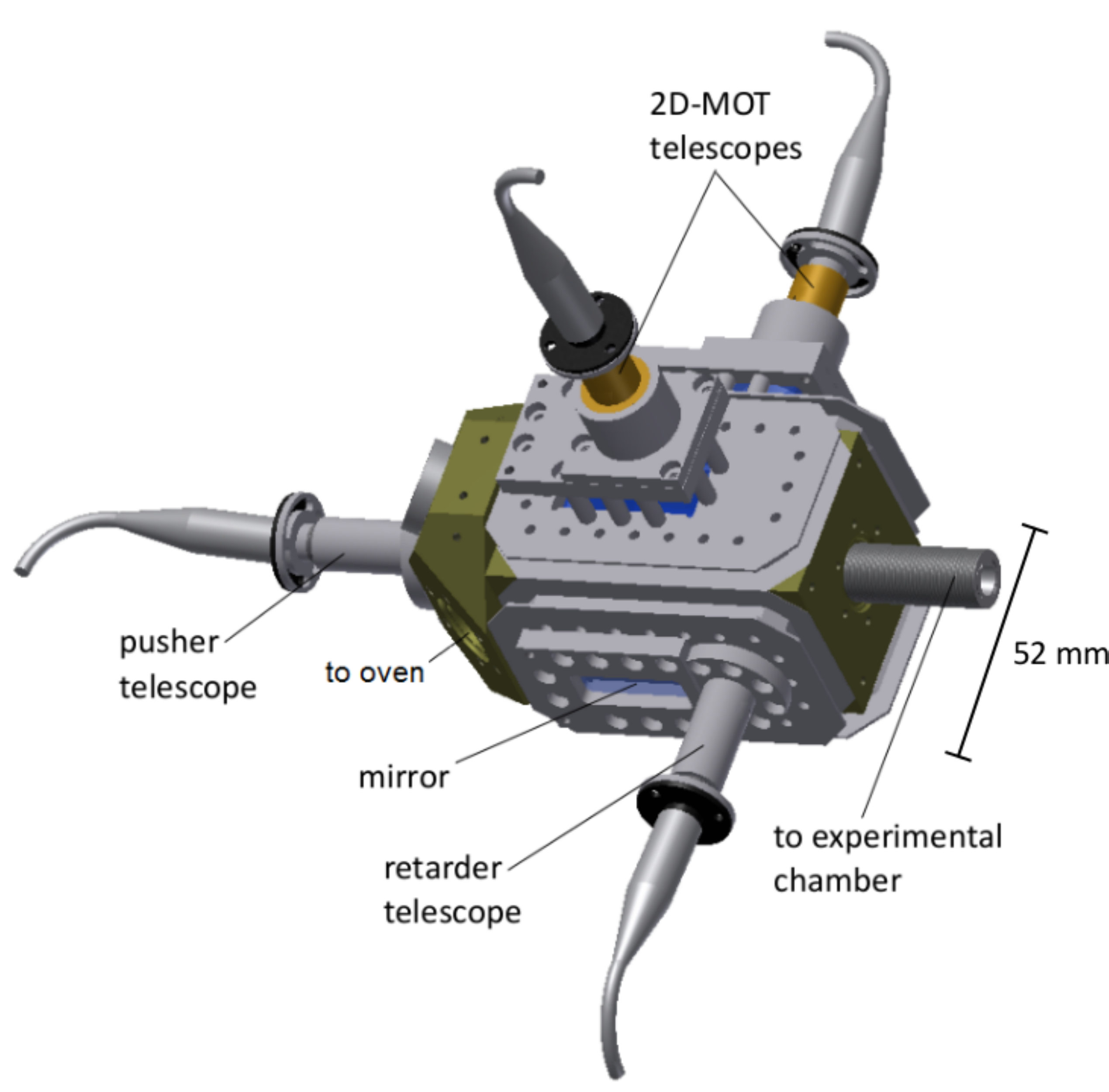}
}
\caption{Source chamber with its ports for optical access.}
\label{sourcechamber}
\end{figure}

\begin{figure}
\resizebox{0.48\textwidth}{!}{%
  \includegraphics{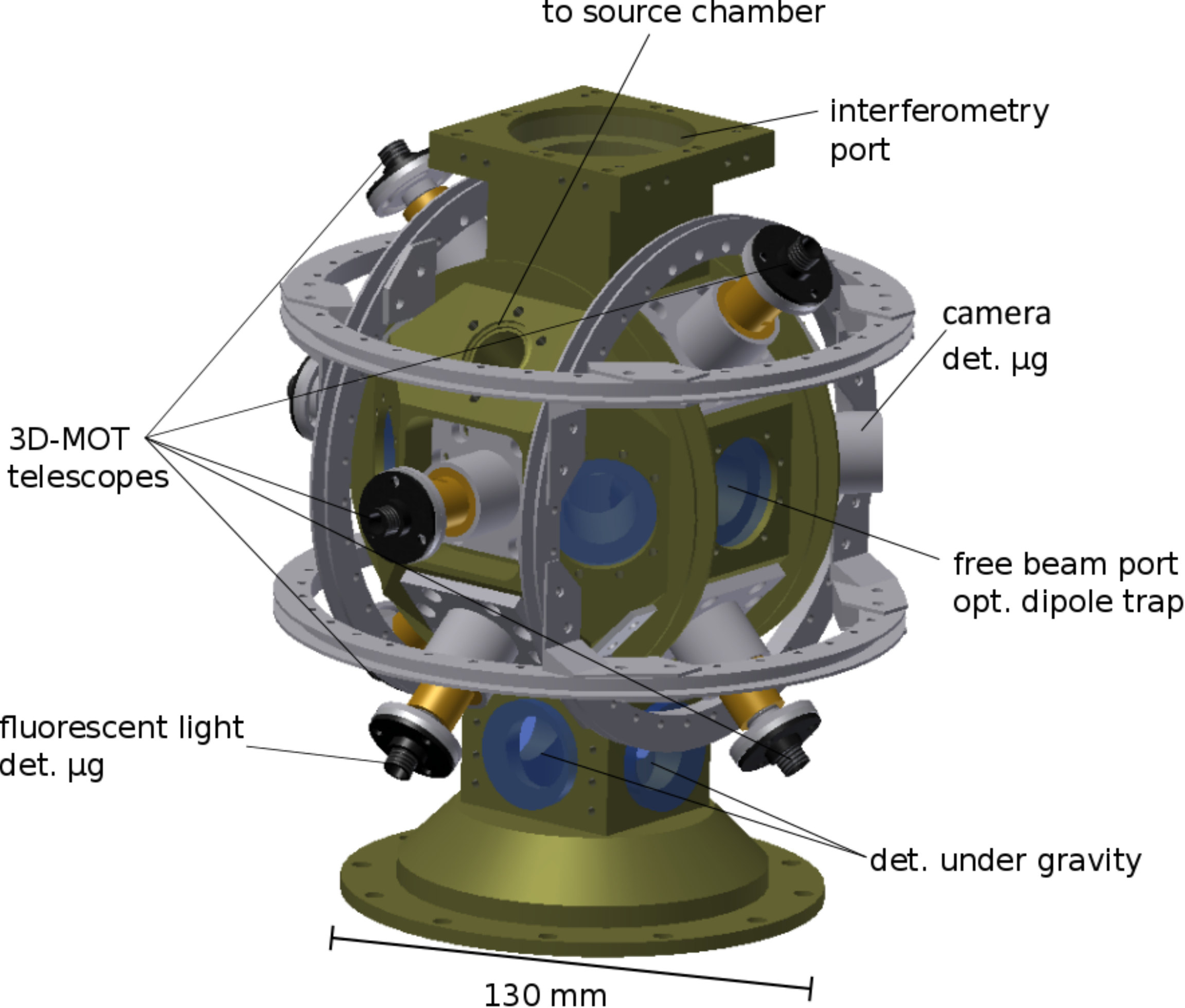}
}
\caption{Experimental chamber with its ports for optical access. Interferometry telescope is not shown.}
\label{experimentalchamber}
\end{figure}

The system of both chambers is $286$\,mm in height.
\begin{figure}
\resizebox{0.48\textwidth}{!}{%
  \includegraphics{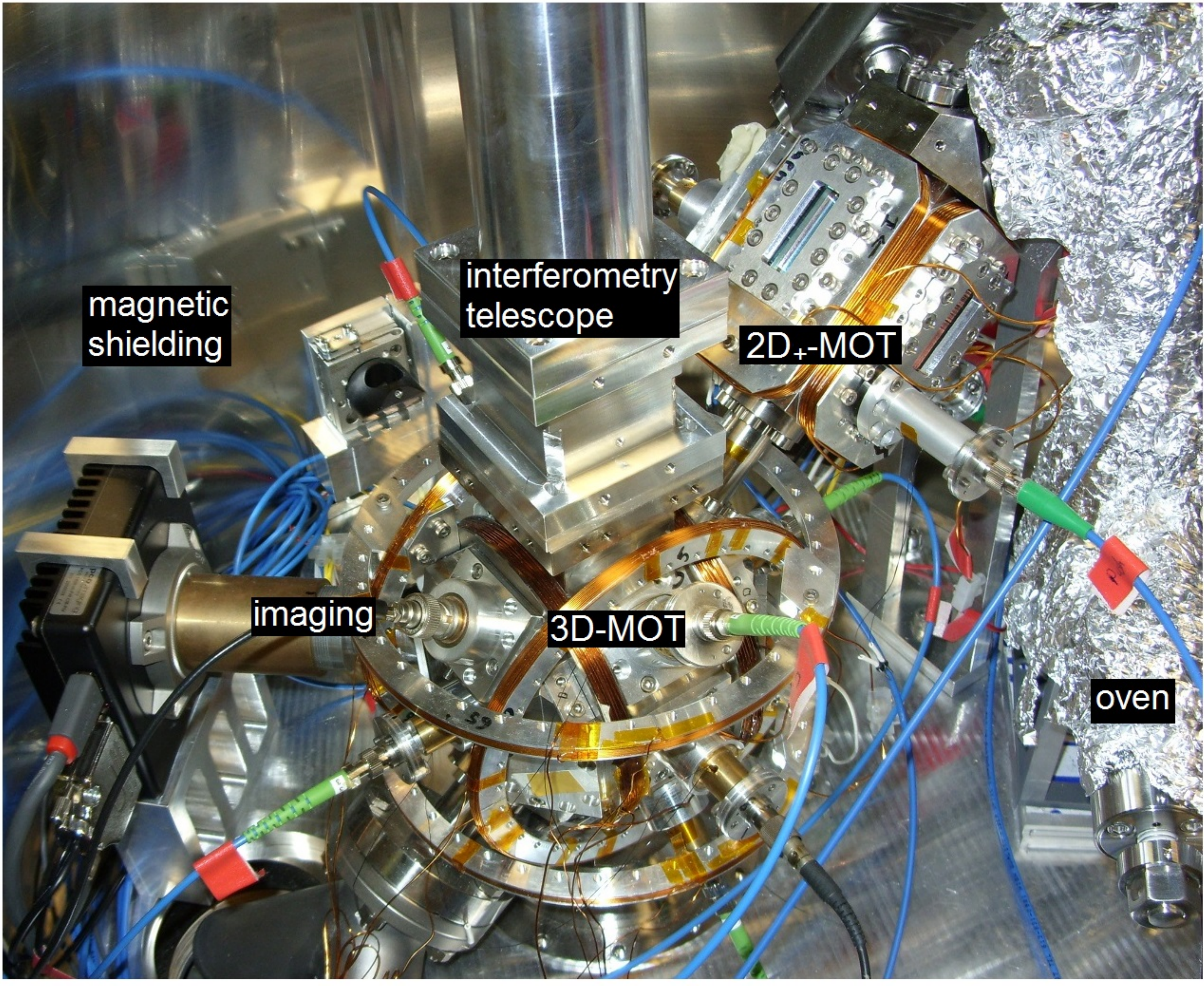}
}
\caption{Vacuum chamber stage with its main components and functions is shown.}
\label{vacuum chamber}
\end{figure}
Each chamber is milled from a block of titanium grade 5 (6Al/4V), which was chosen since it is non-magnetic and has excellent mechanical properties.
The source chamber contains four ports for optical access (see Fig. \ref{sourcechamber}).
Two of them provide the 2D-MOT beams, one the pusher beam and one the retarder beam. The 2D-MOT-beams are retroreflected from a mirror opposite to the respective port. For the distribution of optical power see Table \ref{tab}.\newline The transversely-cooled beam of atoms passes the differential pumping stage and reaches the experimental chamber with the 3D-MOT (see Fig. \ref{vacuum chamber}). The stage consists of a tube of graphite which features a high absorption coefficient for alkali metals. The experimental chamber provides a total of 19 ports for optical access (see Fig. \ref{experimentalchamber}).
The largest port of 2" in diameter on top of the chamber is set for the interferometry beams, which will be vertically aligned and retro-reflected by a mirror at the bottom inside the vacuum chamber. In the center part of the chamber there are six ports for the 3D-MOT beams. Two further ports are for imaging atomic clouds in zero-g and at the bottom section there are four ports for atom imaging under gravity. We use a CCD camera (\textit{PCO}, PCO1400) and a photodiode (\textit{OSI Optoelectronics}, PIN-220D) for fluorescence imaging. All windows are anti-reflection coated and attached to the chamber using indium sealing.
An ion getter pump (\textit{SAES}, NexTorr D100-5) and a chemical getter pump (\textit{SAES}, Capacitorr D400-2) are connected to the experimental chamber (see Fig. \ref{Setup}) to reach ultra-high vacuum conditions in the $10^{-11}$\,mbar regime. Using the differential pumping tube these conditions can be maintained even with a ${\text{2D}}_{+}$-MOT at significantly reduced residual pressure of $>10^{-9}$\,mbar.\newline

\subsection{Two species source / oven}
\label{sec:3}
\begin{figure}
\centering
\resizebox{0.4\textwidth}{!}{
  \includegraphics{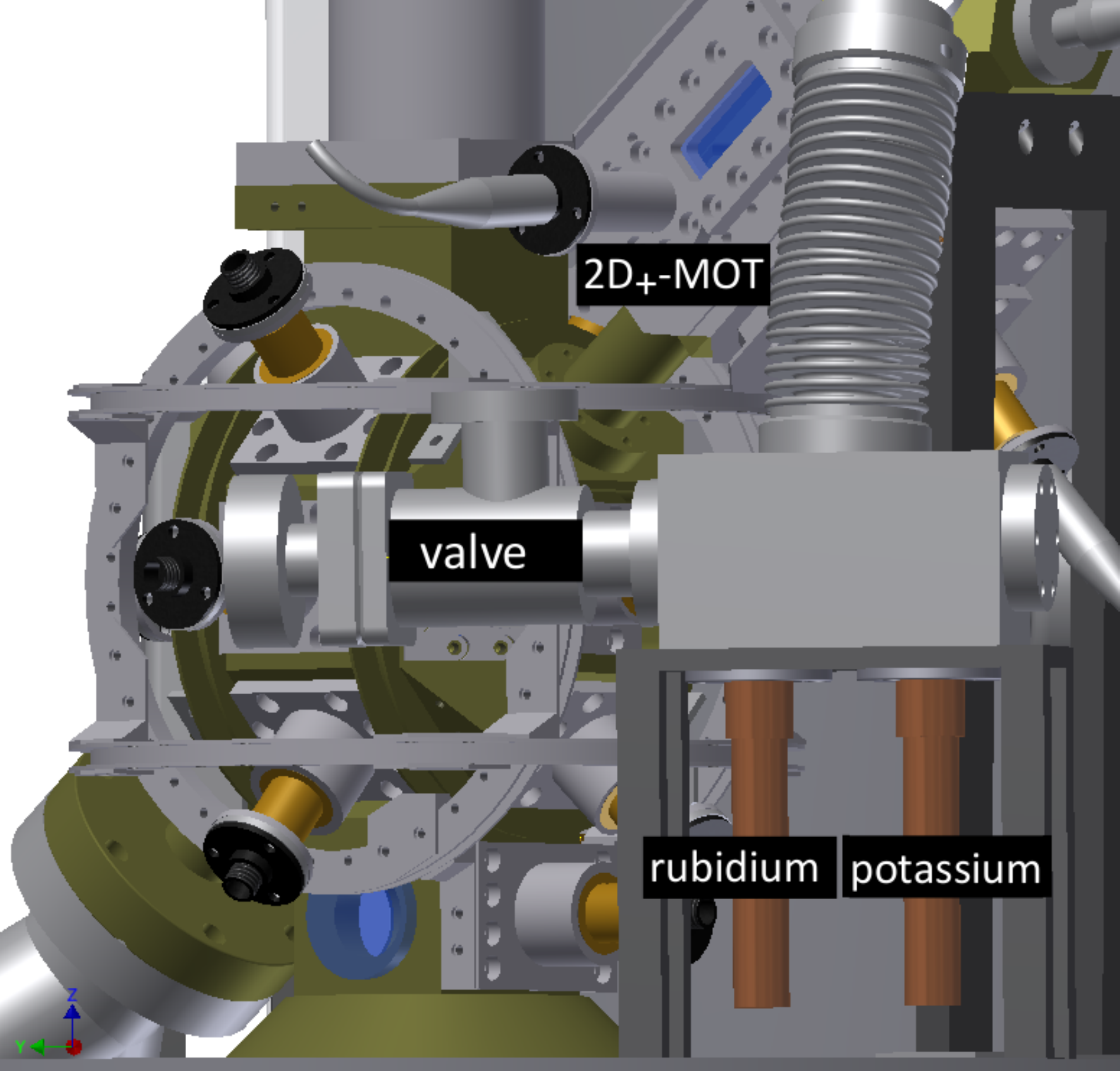}
}
\caption{Source for rubidium and potassium. Heating wires are not shown.}
\label{oven}
\end{figure}

Rubidium and potassium atoms feeding the ${\text{2D}}_{+}$-MOT are provided by two separated reservoirs (see Fig. \ref{oven}) of both elements in their natural abundance. These reservoirs consist of 1\,g break seal glass ampoules housed in a copper tube wrapped with heating wire.
These reservoirs are flanged to a titanium block which is connected to the source chamber by a bellow connector (for mechanical decoupling) and a valve. We keep the rubidium reservoir at a temperature of $64\,{\text{°C}}$ and the one with potassium at $60\,{\text{°C}}$. By measuring the $^{87}{\text{Rb}}$ spectroscopy signal in the source chamber we estimate the $^{87}{\text{Rb}}$ partial pressure to $6.6\times10^{-7}$\,mbar and with a similar measurement we estimate a partial pressure of $1.3\times10^{-9}$\,mbar for $^{39}{\text{K}}$.

\subsection{Magnetic shielding}
\label{sec:4}
External magnetic fields along the drop distance vary by up to 500\,mG (Könemann et al. \citeyear{Koenemann2007}). Our vacuum chamber is thus equipped with a two-layer, cylindric $\muup$-metal shield with a gap of 20\,mm between layers and a thickness of 2\,mm of each layer. It is designed for suppression of external fields by a factor of $>$100 in all directions.
This is particularly important for the final stages of the laser cooling (molasses cooling) we plan, which requires suppression of external magnetic fields to the low mG level. \newline
Furthermore, future interferometry measurements to be performed with this setup will also be very sensitive to resi\-dual magnetic fields. A detailed discussion of systematic effects due to residual magnetic fields in interferometry experiments can be found in Aguilera et al. (\citeyear{Aguilera2014}).

\begin{figure}
\resizebox{0.49\textwidth}{!}{
  \includegraphics{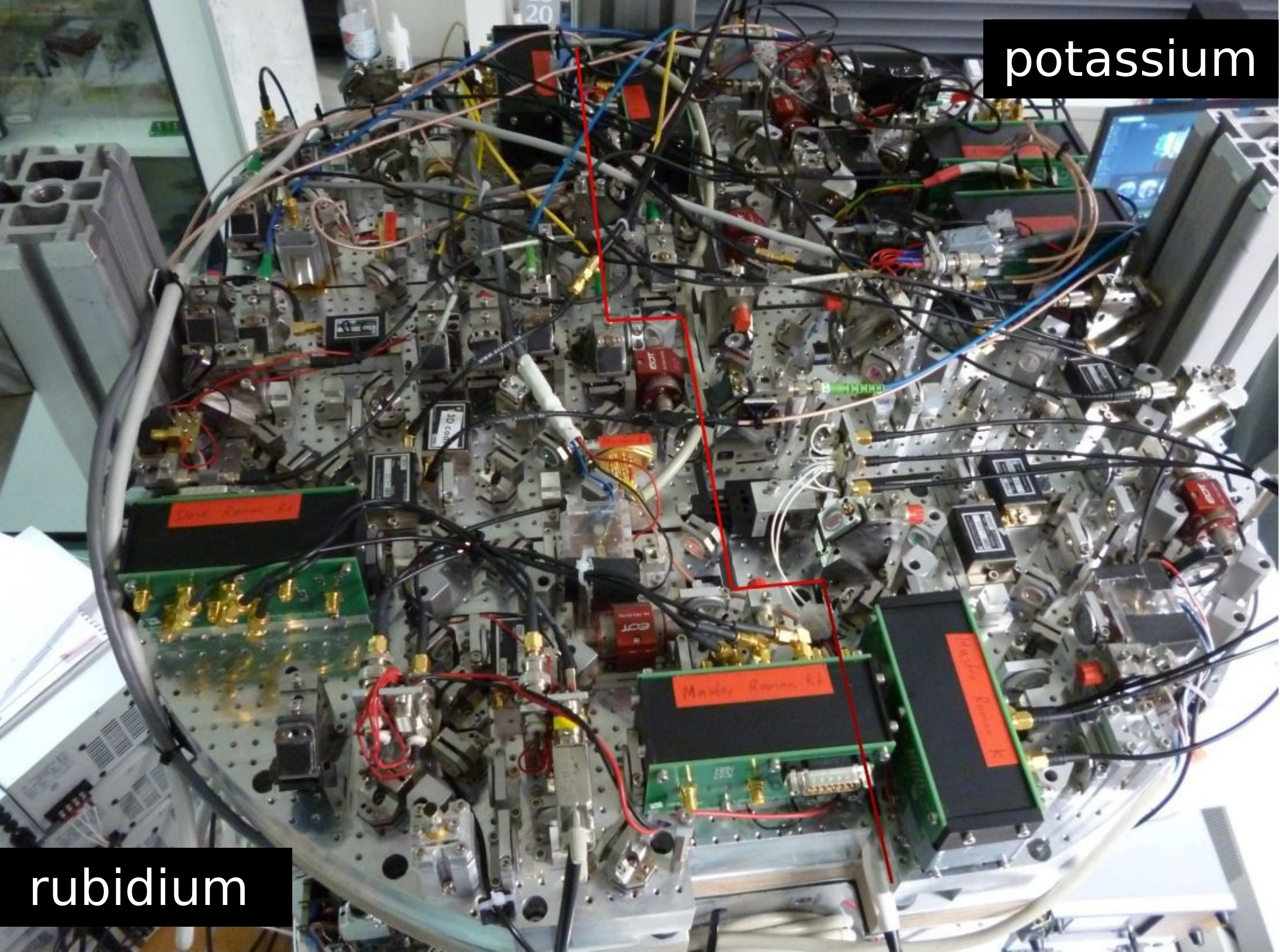}
}
\caption{Top view of the laser systems for $^{87}{\text{Rb}}$ (left) and $^{39}{\text{K}}$ (right).}
\label{TopLasersystem}       
\end{figure}
\section{Laser Systems}
\label{sec:5}
\begin{figure*}[tb]
    \centering
		\resizebox{1.0\textwidth}{!}{%
    \includegraphics{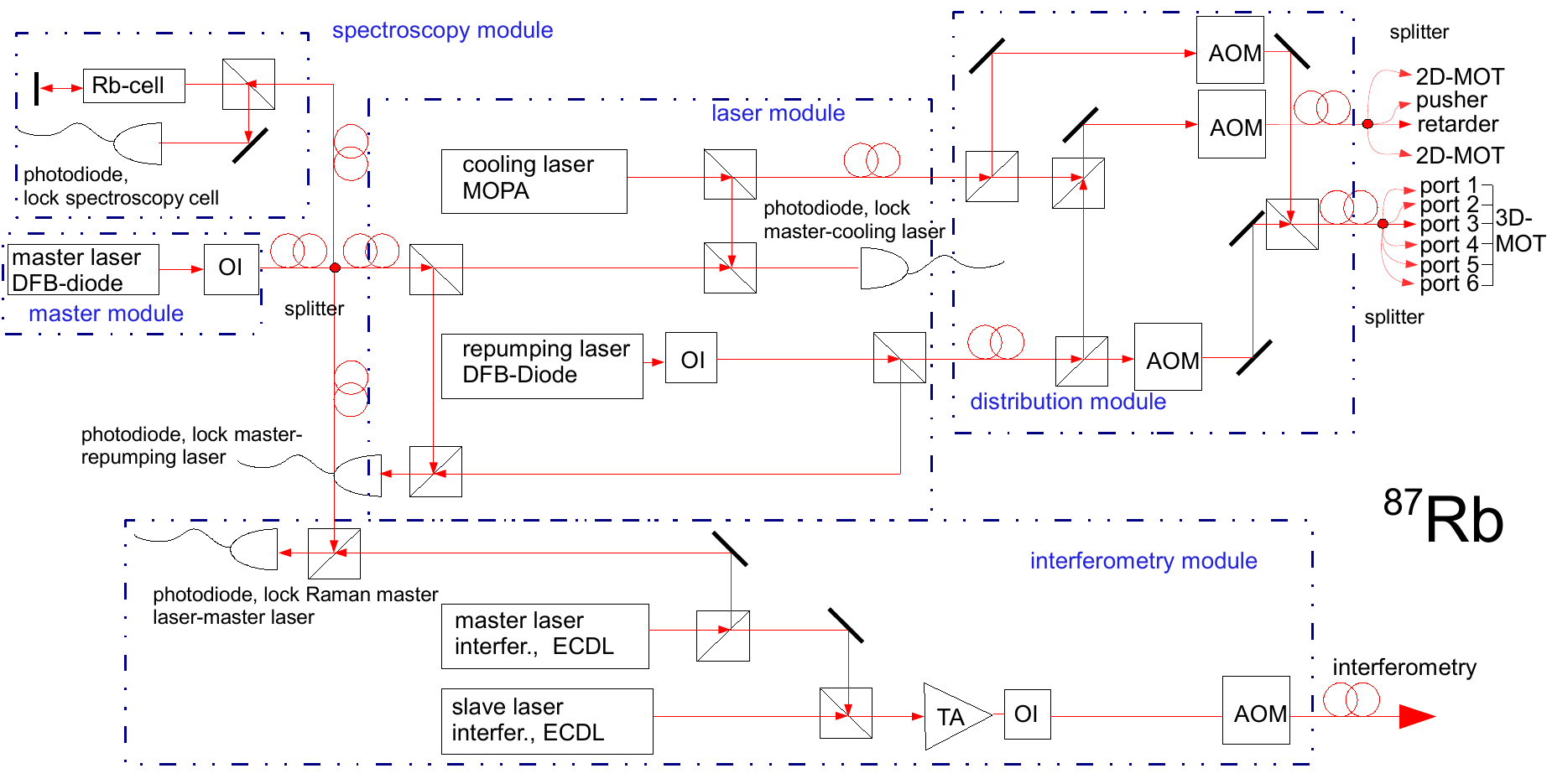}
		}
    \caption{Sketch of rubidium laser system setup. OI: Optical Isolator, TA: Tapered Amplifier}
    \label{setup1}
\end{figure*}

Figure \ref{TopLasersystem} shows the laser systems for cooling $^{87}{\text{Rb}}$ and $^{39}{\text{K}}$ atoms.
The rubidium part at 780\,nm takes the left half of the platform with 0.7\,m in diameter and the potassium part at 767\,nm the right half. Both parts can be subdivided into four modules respectively: a laser module, a distribution module, a master laser module and an interferometry module. Every module consists of a 40\,mm thick base plate made of aluminum to provide sufficient bending stiffness. In oder to reduce its weight a 30\,mm deep polygone structure is milled into the bottom side of the base plate.\newline
The reference laser (master laser) for the rubidium laser system is a DFB diode (\textit{Eagleyard},EYP-DFB-0780-00080-1500-TOC03-0000), which is stabilized onto the $\left|^{2}S_{1/2},F=3\right\rangle\rightarrow\left|^{2}P_{3/2},CO:F'=3,4\right\rangle$-transition of $^{85}{\text{Rb}}$ using Doppler-free saturation spectroscopy (see Fig. \ref{setup1}). The spectroscopy cell has the dimensions $103$ mm $\times$ $25$\,mm (length $\times$ diameter). The spectroscopy module is fiber coupled and due to its size its placed below the vacuum chamber stage (see Fig. \ref{Setup}).

The light of the master laser is divided and distributed to all other laser modules by optical fibers. There it is overlapped with light from the respective other laser on a fast photodiode for generation of a beat-note and subsequent offset locking of the respective laser. For the MOT-systems cooling and repumping light is needed. For the cooling light we use a master-oscillator power amplifier (MOPA) developed at the \textit{Ferdinand-Braun-Institut} (\textit{FBH}) in Berlin (Schiemangk et al. \citeyear{Schiemangk2015}). The MOPA used here is integrated on a micro-optical bench (MIOB) and has an output power of 1450\,mW, an intrinsic line\-width of 35\,kHz and a FWHM linewidth of $1.5$\,MHz. The repumping light is provided by a DFB-diode of the same type as the master laser.

\begin{figure}
\resizebox{0.46\textwidth}{!}{
  \includegraphics{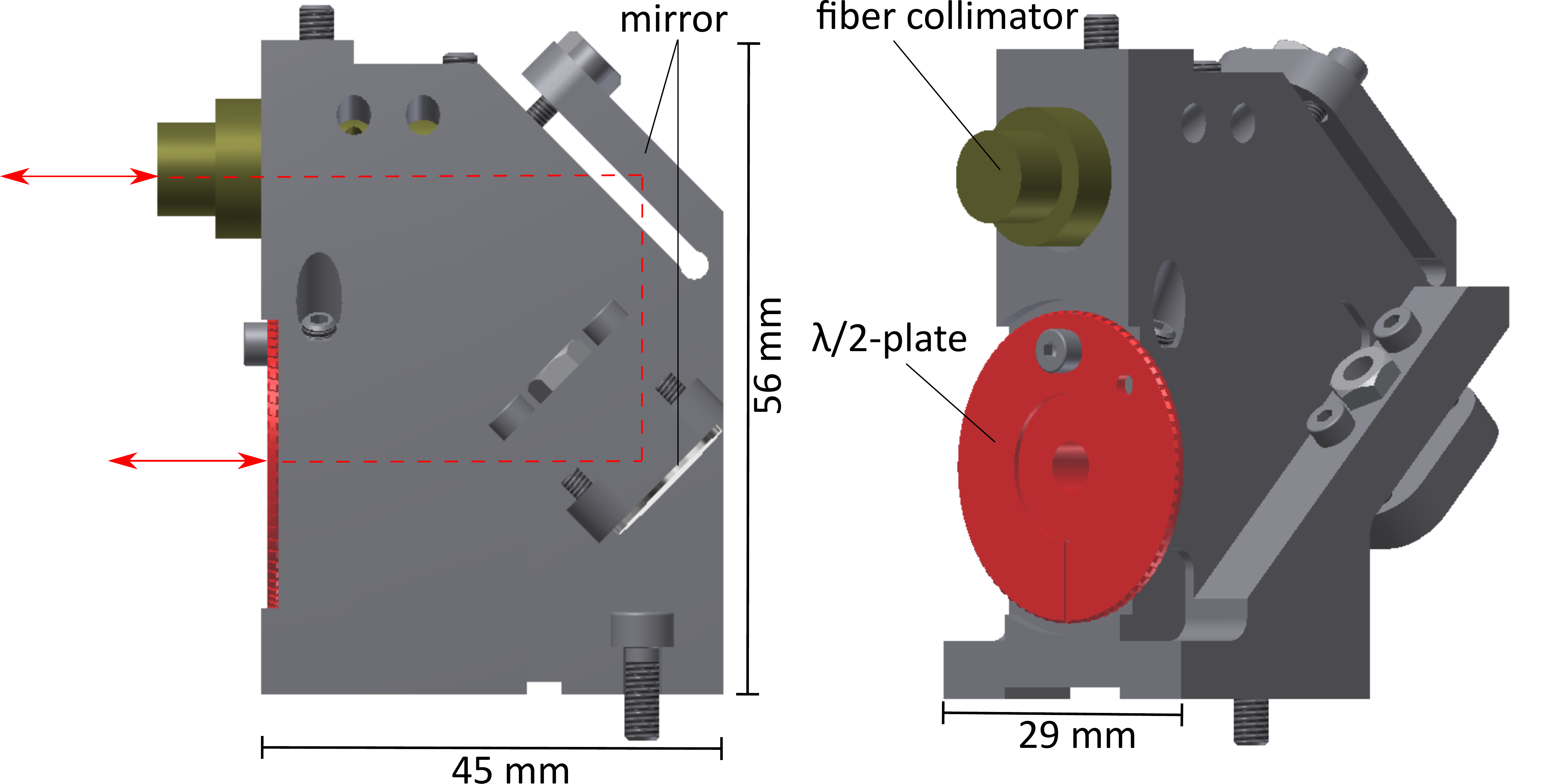}
}
\caption{In the laser systems miniaturized optical periscope-like mounts are used for compact and stable fiber coupling.}
\label{periscope}
\end{figure}
\begin{figure}
\resizebox{0.48\textwidth}{!}{
  \includegraphics{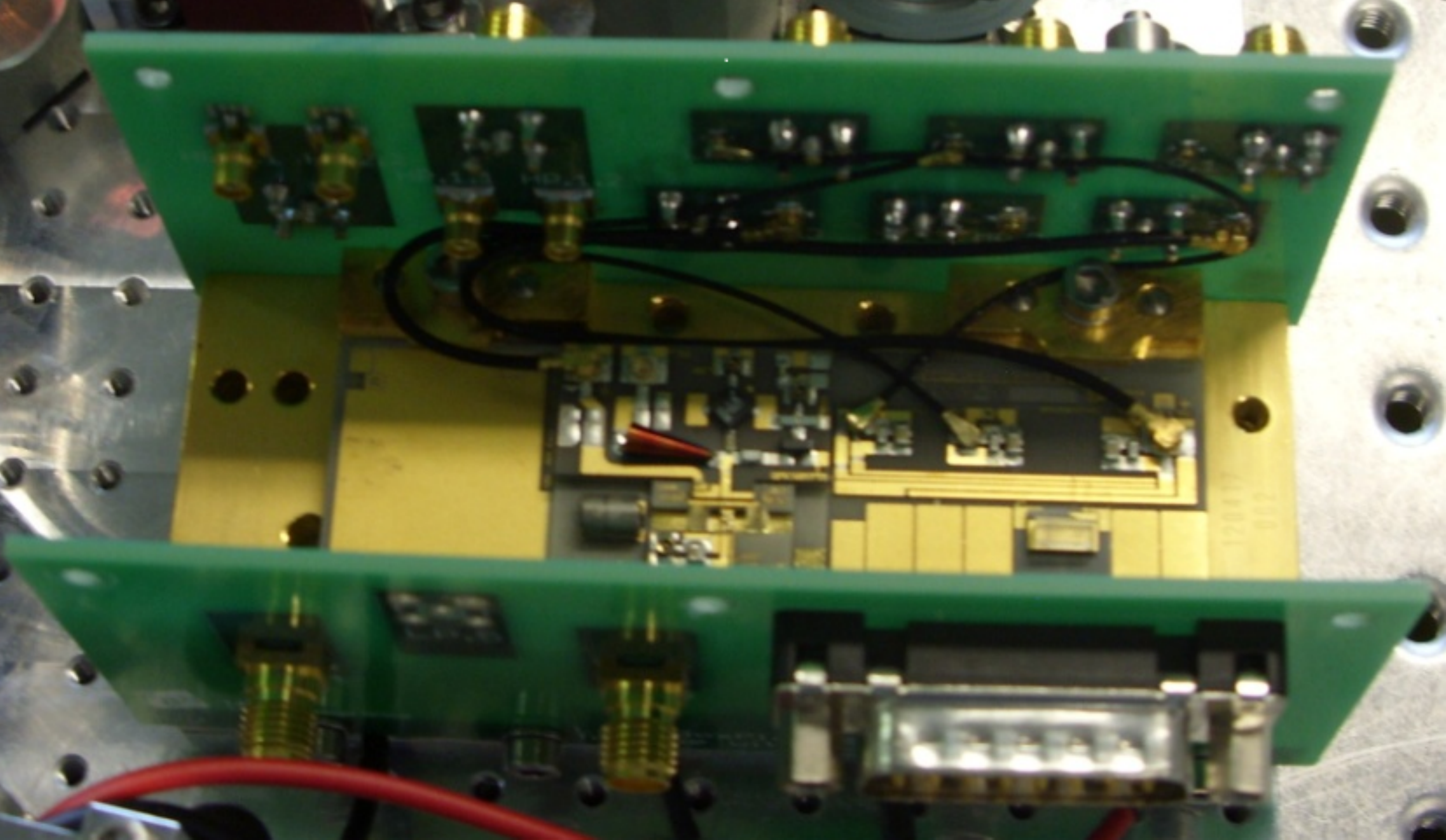}
}
\caption{Photo shows a micro-integrated extended cavity diode laser module for Rb-interferometry with integrated optical isolator on a micro-optical bench consisting of aluminum nitride ceramics. A part of the housing has been removed for reasons of clarity.}
\label{ECDL}
\end{figure}

Repumping and cooling light are coupled into a fiber and sent to the distribution module. For compact and stable fiber coupling we use a miniaturized optical periscope-like mount, which features two adjustable mirrors, a half-wave-plate and a fiber collimator (see Fig. \ref{periscope}).
On the distribution module both laser beams are split, overlapped along two paths and sent to the ${\text{2D}}_{+}$-MOT and 3D-MOT fiber couplers. For splitting the light and guiding it to the different ports at the ${\text{2D}}_{+}$- and 3D-chambers fused fiber splitters (${\text{2D}}_{+}$-MOT: \textit{OZ Optics}, 3D-MOT: \textit{Evanescent Optics}) are used with a splitting ratio as shown in Table \ref{tab}. Finally, the distribution unit also provides one port to couple light for fluorescence detection, which is not shown in Figure \ref{setup1}. Acousto-optical modulators (\textit{Crystal Technology}, 3080-125) are used as fast light switches in the experimental sequence and additionally for implementing laser frequency detunings. \newline The setup of the interferometry module to generate the beamsplitter and mirror pulses can be seen in Figure \ref{setup1}. Here, we use two extended cavity diode lasers (ECDLs) to drive a two-photon Raman transition between the hyperfine levels of the Rb ground state with a frequency difference of 6.8\,GHz. The ECDLs are also provided by the \textit{FBH}, integrated on micro-optical benches similar to the MOPA (see Fig. \ref{ECDL}) and feature an intrinsic linewidth of $0.4$\,kHz and a FWHM linewidth  of $135$\,kHz (Luvsandamdin et al. \citeyear{Luvsandamdin2014}). This narrow linewidth is favorable for phase locking of the lasers. The feedback cavity is implemented using a tempera\-ture-stabilized volume holographic Bragg grating (VHBG) and the front facet of the ridge wave\-guide amplifier chip (RW).
One ECDL serves as the Raman master laser which is offset locked to the master laser of the rubidium system. The other ECDL is the so called slave laser, that will be phase locked to the Raman master laser.
The low-phase-noise reference signal will be generated similarly as described in Gouet et al. (\citeyear{Gouet2008}). In particular we have acquired a frequency chain from \textit{Rupptronik} providing a 6.9\,GHz signal, referenced to an ultra-low-phase-noise 100\,MHz quartz oscillator. At 10\,Hz offset from the 6.9\,GHz carrier this system allows to reduce phase noise to $< 85$\,dBc/Hz.
\begin{figure*}[tb]
    \centering
		\resizebox{1.0\textwidth}{!}{%
    \includegraphics{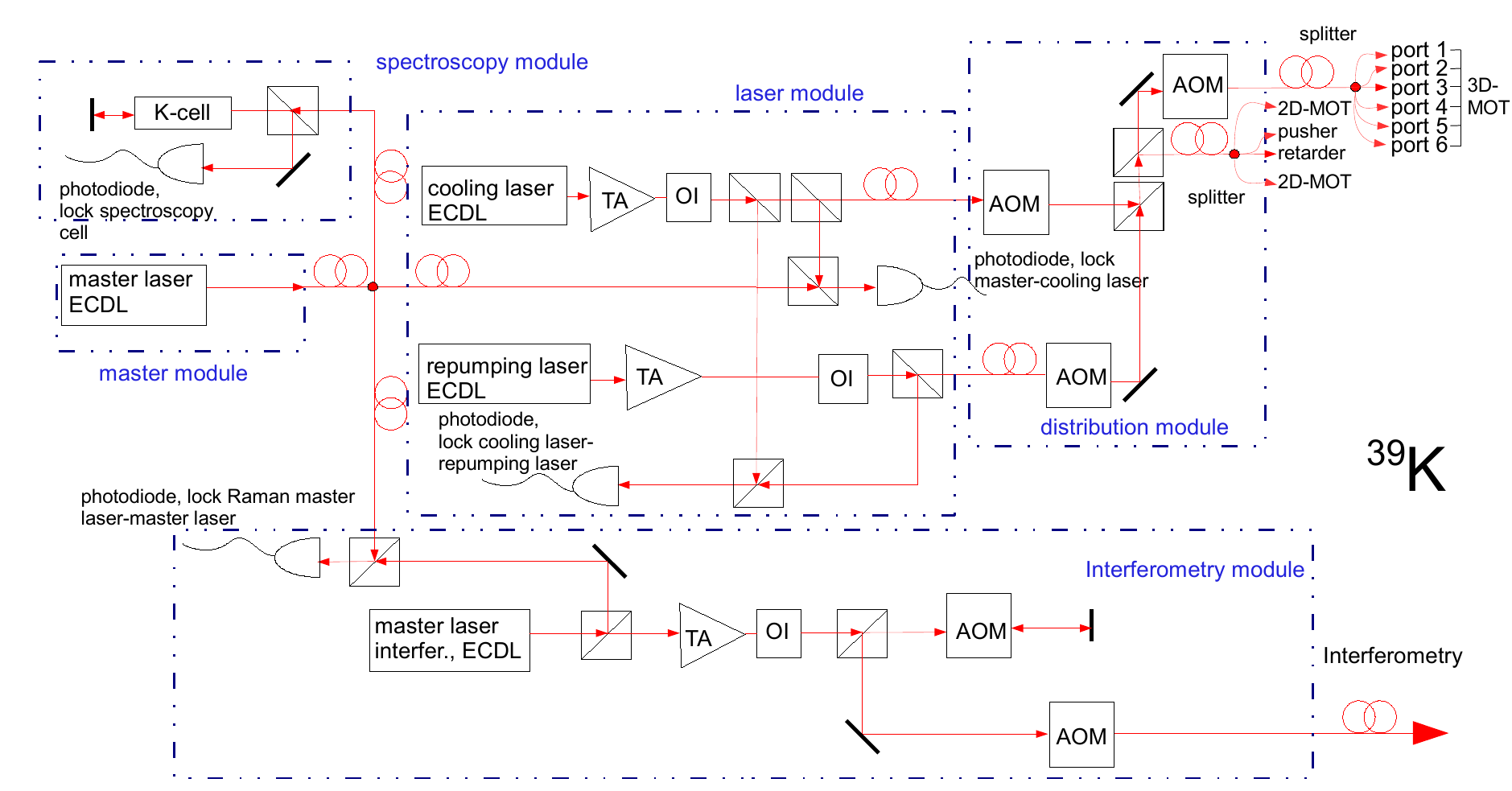}
		}
    \caption{Sketch of potassium laser system setup. OI: Optical Isolator, TA: Tapered Amplifier}
    \label{setup2}
\end{figure*}

The light of the Raman slave laser and the Raman master laser is combined and amplified by a tapered amplifier (\textit{Eagleyard}, EYP-TPA-0780-02000-4006-CMT04-0000). It passes an acousto-optical modulator (AOM) to switch the mirror and beamsplitter pulses for the Raman interferometry.

On all laser modules optical isolators ensure protection of the respective laser diodes from undesired back reflections. Single stage macroscopic isolators from \textit{EOT} (04-780-00766) with 30\,dB isolation are inserted behind the tapered amplifier and the DFB-diodes, while those lasers which are mounted on MIOBs are already equipped with miniaturized optical isolators with 30\,dB / 60\,dB  isolation.

The structure of the potassium system (see Fig. \ref{setup2}) is similar to that of the rubidium system. The main difference is the use of narrow linewidth ECDLs not only for interferometry but for cooling, repumping and spectroscopy as well. This becomes important due to the small hyperfine level splittings of only few MHz. The ECDLs however only deliver about 20\,mW of optical power. Therefore we use tapered amplifiers (\textit{Eagleyard}, EYP-TPA-0765-01500-3006-CMT03-0000) to amplify the cooling and repumping light. For $^{39}{\text{K}}$ the repumping power has to be similar to the cooling power because of an inefficient cooling cycle (Landini et al. \citeyear{Landini2011}).\newline The potassium master laser is locked to the crossover resonance $\left|^{2}S_{1/2},CO:F=1,2\right\rangle \rightarrow \left|^{2}P_{3/2},F'=1,2,3\right\rangle$ of $^{39}{\text{K}}$. A potassium-filled vapor cell with the dimensions $25$\,mm $\times$  $25$\,mm (length $\times$ diameter) is heated to a temperature of about $65$\,°C in order to increase the low vapor pressure of $^{39}{\text{K}}$ at room temperature. The spectroscopy module for potassium can also be found below the vacuum chamber stage (see Fig. \ref{Setup}).

The scheme for frequency-stabilization is realized as follows: the repumping laser is locked onto the cooling laser which in turn is locked onto the master laser, thus avoiding a beat-note frequency close to zero. The distribution for the ${\text{2D}}_{+}$-MOT and 3D-MOT differs from the scheme applied in the rubidium system such that one beam splitter cube splits the combined cooling and repumping light with similar power for the two MOT-systems in a ratio of 50\,:\,50 to handle the inefficient cooling cycle (Landini et al. \citeyear{Landini2011}).\newline
Due to the smaller hyperfine transition frequency for the $\left|^{2}S_{1/2},F=1\right\rangle \rightarrow \left|^{2}S_{1/2},F=2\right\rangle$-transition for $^{39}{\text{K}}$ of 461.7\ MHz we can create two frequency-shifted Raman beams from only one laser by using a double-pass AOM (\textit{Crystal Technology}, 3200-125). This is followed by another AOM for switching the interferometer pulses. For the RF source of the double-pass AOM the 100\,MHz low-phase-noise reference will be doubled in frequency and mixed with a low-phase-noise DDS signal.
\begin{table}
\caption{Laser power for ${\text{2D}}_{+}$- and 3D-MOT for rubidium, measured at the vacuum chamber windows.}
\label{tab}
\begin{tabular}{lll}
\hline\noalign{\smallskip}
\textbf{rubidium} & cooler & repumper  \\
\noalign{\smallskip}\hline\noalign{\smallskip}
2D telesc.1,2 & 33\,mW & 1.3\,mW  \\
pusher & 3.8\,mW & 0.3\,mW \\
retarder & 0.64\,mW & 0.06\,mW \\
3D telesc.1-6 & 5.2\,mW & 0.8\,mW \\
\noalign{\smallskip}\hline
\end{tabular}
\end{table}

\subsection{Electronics and Power Management}\label{sec:6}
\begin{figure}
\resizebox{0.48\textwidth}{!}{%
  \includegraphics{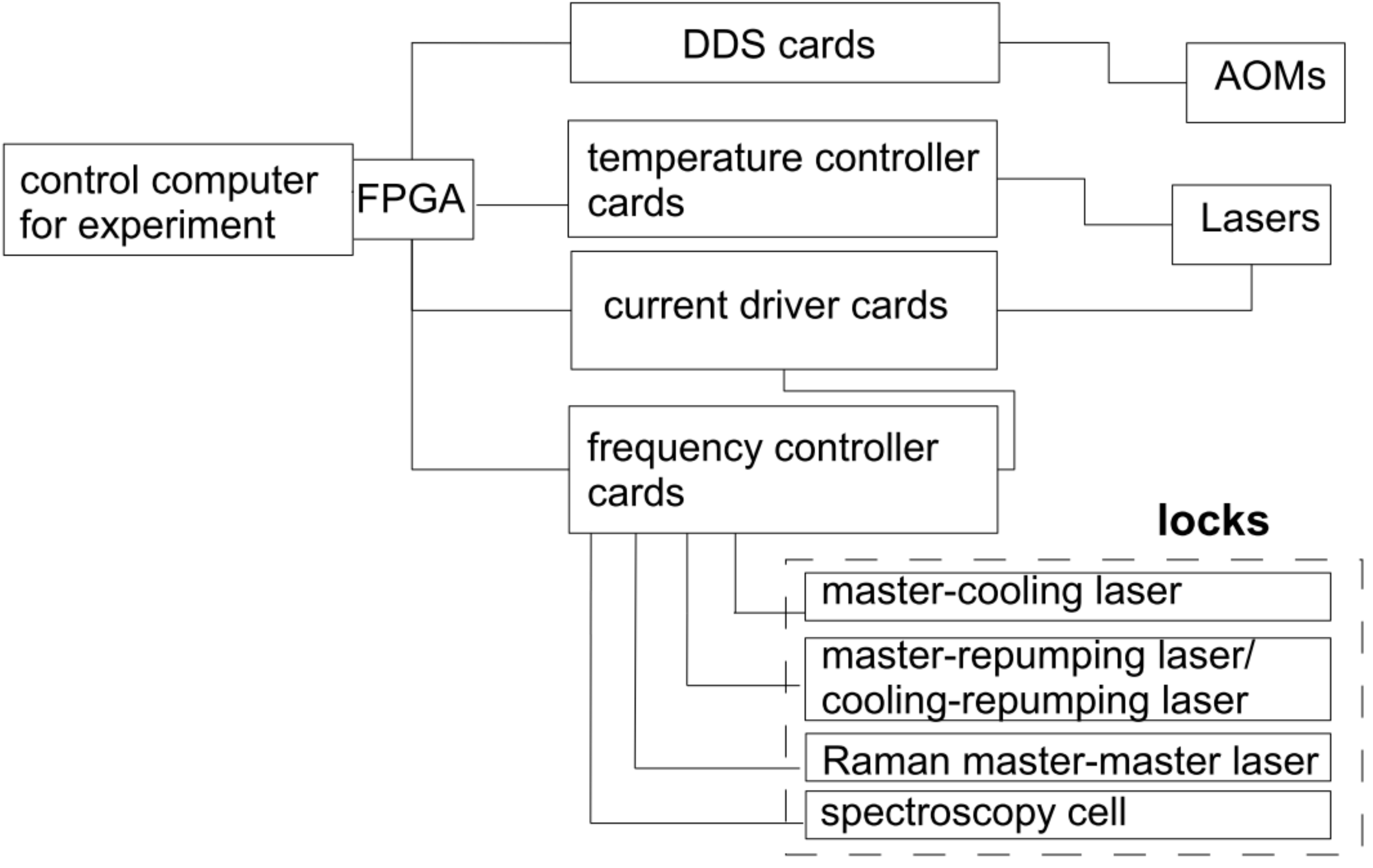}
}
\caption{Sketch of stack communication for $^{87}{\text{Rb}}$ and $^{39}{\text{K}}$ laser systems.}
\label{comm}
\end{figure}

The electronics used in the setup like temperature controller, current driver and frequency controller for both laser systems are custom-designed as stacks of electronic boards of $100$\,mm $\times$  $100$\,mm using a common bus communication (see Fig. \ref{comm}).

A PXI system (\textit{National Instruments}, NI PXI-8196) acts as control computer in the capsule and a real-time operating system for remote control is installed on it. The experimental sequence is written into an FPGA (field programmable gate array) chip (\textit{National Instruments}, NI PXI-7833R) communicating with the stack.\newline
The frequency controller cards use the various beat-note signals provided from the laser systems (Figure \ref{setup1},\ref{setup2}) for offset locking the respective laser frequencies in reference to the spectroscopy laser. The DDS cards generate the radio frequencies that are fed to 1\,W power amplifiers and drive the AOMs.\newline
The battery platform at the bottom of the system provides the electric power during the drop and is buffered by power supplies in the lab. It delivers the power for the rubidium and potassium laser systems, the magnetic coils and the peripheral electronics such as camera and fans. Under full load (lasers on, magnetic coils on, vacuum pumps on, etc.) the system needs 535\,W and can be powered for up to 3.6 hours. The required power is divided onto different battery pack configurations consisting of 3.2V-LiFePO4-cells.

\section{Performance of the MOT-system}
\label{sec:7}
\begin{figure*}[tb]
\centering
\resizebox{1.0\textwidth}{!}{%
    \includegraphics{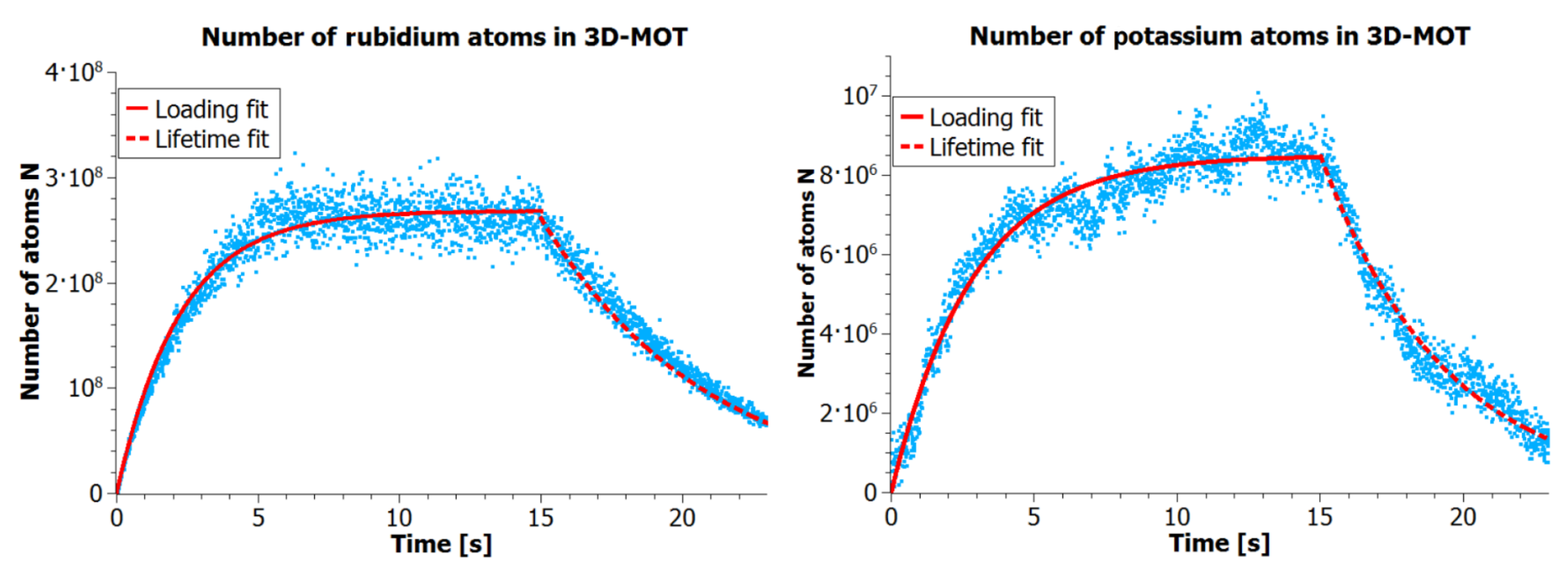}
    }
    \caption{Loading rate and lifetime of $^{87}{\text{Rb}}$ and $^{39}{\text{K}}$ in 3D-MOT in the laboratory.}
    \label{MOT}
\end{figure*}

Figure \ref{MOT} presents the 3D-MOT loading for $^{87}{\text{Rb}}$ and $^{39}{\text{K}}$ in the laboratory. The 3D-MOT is loaded for 15\,s and then the 2D-MOT is switched off. Here the fit $N(t)=\frac{L}{\Gamma}\left({1-e^{-\Gamma t}}\right)$ is used to characterize the loading rates. This fit includes losses due to collisions with non-rubidium/potas\-sium atoms by the factor $\Gamma$. Density-dependent losses are neglected and $L$ is the loading rate from the 2D-MOT here. The atom number was measured by fluorescence imaging and is plotted as a function of time. We achieve a value of $N_{\text{0R}}=2.7\times10^{8}$ for the maximum rubidium atom number and $N_{\text{0K}}=8.5\times10^{6}$ for the maximum potassium atom number.
The initial loading rates are $L_{\text{0R}}=9.5\times10^{7}\,{\text{s}}^{-1}$ and $L_{\text{0K}}=3.0\times10^{6}\,{\text{s}}^{-1}$. 
We also determined the lifetime $\tau_{1}$ in the 3D-MOT as can be seen in Figure \ref{MOT}. With $N=N_{0}e^{-t/\tau_{1}}$ we obtain $\tau_{\text{1R}}=5.6$\,s and $\tau_{\text{1K}}=4.3$\,s. Based on the loading and lifetime plots the vacuum pressure within the experimental chamber can be estimated to be $p=7\times10^{-11}$mbar. A first temperature measurement of $145\,\muup$K is obtained from an analysis of the rubidium ensemble's expansion at different time of flights.
The comparably small potassium atom number can main\-ly be explained by the rather small cooling and repumping light power, e.g. 2D-cooling/repumping power per telescope is 17.6\,mW/13.2\,mW and 3D-cooling/repumping power per telescope is 5.6\,mW/3.7\,mW. Changes of the laser system setup to increase the optical power are in progress. The oscillation in the atom number of potassium (see Fig. \ref{MOT}) can be explained by fluctuations of laser power. Due to the comparably small atom number the fluctuations are visible. Since loading can be done previous to the capsule release within arbitrarily long time, the system was not optimized for fastest loading rate.\newline
Finally, we have been able to operate both MOTs simultaneously as required for implementation of the next steps in our experiment, as discussed below.

\section{Dipole Trap}
\label{sec:8}
\subsection{Thulium Fiber laser}
\label{sec:9}
For further cooling down to sub-$\muup$K temperature we plan to employ the process of evaporative cooling in an optical dipole trap.
The laser used for the dipole trap is a high-power linearly-polarized thulium fiber laser at 1950\,nm wavelength with 26\,W cw power and a FWHM linewidth of 0.16\,nm. Due to the $\delta^{-2}$ dependency of the heating rate, where $\delta$ is the detuning to the linewidth of the first atomic transition of $^{87}{\text{Rb}}$, the photon scattering rate is comparatively small. However, since polarizability scales as $\delta^{-1}$ a high laser power is needed for trapping the atoms. For $^{39}{\text{K}}$ these effects seem to have a similar dependency but have still to be analyzed in more detail.\newline The fiber laser was built specifically for the use in the drop tower by the \textit{Laser Zentrum Hannover e.V.}, in a compact and robust design to withstand the deceleration of the drop capsule. It is a two-stage system with a fiber laser oscillator followed by a fiber amplifier. The power amplifier stage of the system is characterized by a slope efficiency of $56\,\%$. Both stages are driven by separate current drivers and temperature controllers, which stabilize the temperature to 25\,°C. In the lab the laser system is connected to a cooling water circuit with a water temperature of 21\,°C, sustained by a chiller (\textit{Thermo Scientific}, ThermoFlex 1400). This is necessary due to the heat dissipation mainly by the pump laser diodes on the power stage. During operation in free fall no water cooling is available, thus a phase-change material is implemented which acts as a heat reservoir by changing from solid to liquid. A surge tank compensates its volume expansion of about $10\,\%$.

Our setup includes the components for a weak hybrid trap which is a combination of a single-beam dipole trap and an additional weak magnetic quadrupole field to increase the atomic density in the trap. The weak hybrid trap has been demonstrated in a setup at the \textit{Institut für Quantenoptik} in Hannover (Zaiser et al. \citeyear{Zaiser2011}). The advantage of this technique compared to two-beam setups (Han et al. \citeyear{Han2001}; Clément et al. \citeyear{Clement2009}) is the comparatively easy and robust setup, which is important for microgravity experiments.

As starting conditions for the hybrid trap for rubidium we assume a compressed $^{87}{\text{Rb}}$ MOT with a density of $1\times10^{11}$ atoms/{cm}$^3$ and an optical power for the dipole trap of 10\,W. We choose a beam focus of $\omega_{0}=50\,\muup{\text{m}}$. With a laser power of 10\,W for the trapping beam this will create a trap with radial/axial trap frequencies of $\omega_{\text{rad}}=2\pi \times 800\,{\text{Hz}}$ / $\omega_{\text{ax}}=2\pi \times 7\,{\text{Hz}}$, with a trap depth of $U_{0}=k_{B}\times 190\,\muup{\text{K}}$. With this we plan to trap up to $10^6$ atoms as achieved in Zaiser et al. (\citeyear{Zaiser2011}). \newline
\subsection{Fiber laser periphery}
\label{sec:10}
\begin{figure}
\centering
\resizebox{0.4\textwidth}{!}{
  \includegraphics{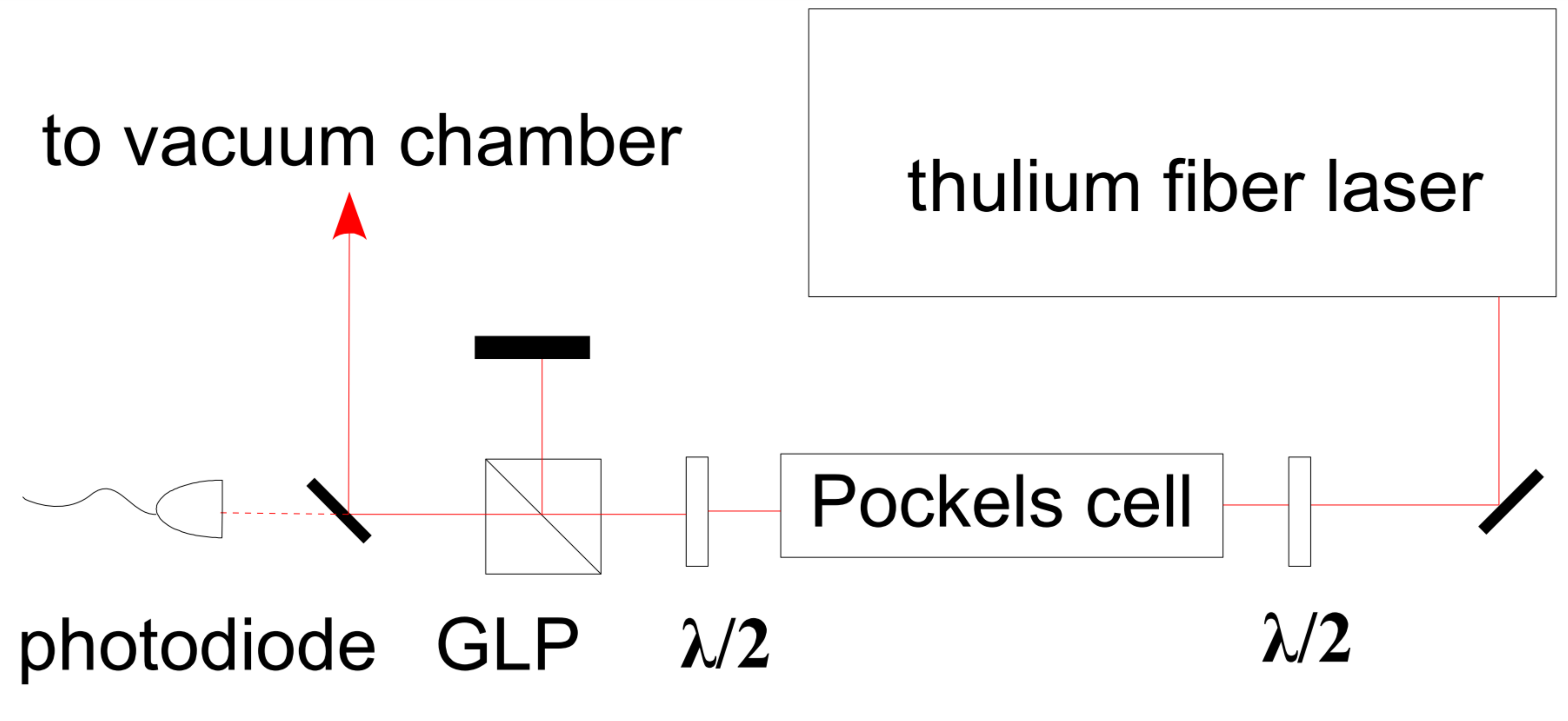}
}
\caption{Sketch of Pockels-cell setup.}
\label{Pock}
\end{figure}

The thulium fiber laser provides a free beam at its output, which passes a $\lambda/2$-waveplate, a Pockels cell (driven by a high-voltage amplifier which generates 7\,kV maximum), another $\lambda/2$-waveplate and a Glan-Laser polarizer (GLP) to regulate its power (see Fig. \ref{Pock}). Precise control of laser power is important for the process of evaporative cooling, which we plan to employ for cooling the atoms to sub-$\muup$K temperature (Zaiser et al. \citeyear{Zaiser2011}). The transmission of the Pockels cell (\textit{Linos}, LM7IR) as function of the applied voltage is given by a cosine wave-form. In order to facilitate the implementation of a feedback control this transmission function is linearized using its inverse function. This feature is currently being implemented in a closed-loop control system. Additionally, the system will reduce power fluctuations of the dipole laser power to below 0.1\,mW, because larger fluctuations lead to heating and potentially loss of atoms.

\begin{figure}
\centering
\resizebox{0.48\textwidth}{!}{
  \includegraphics{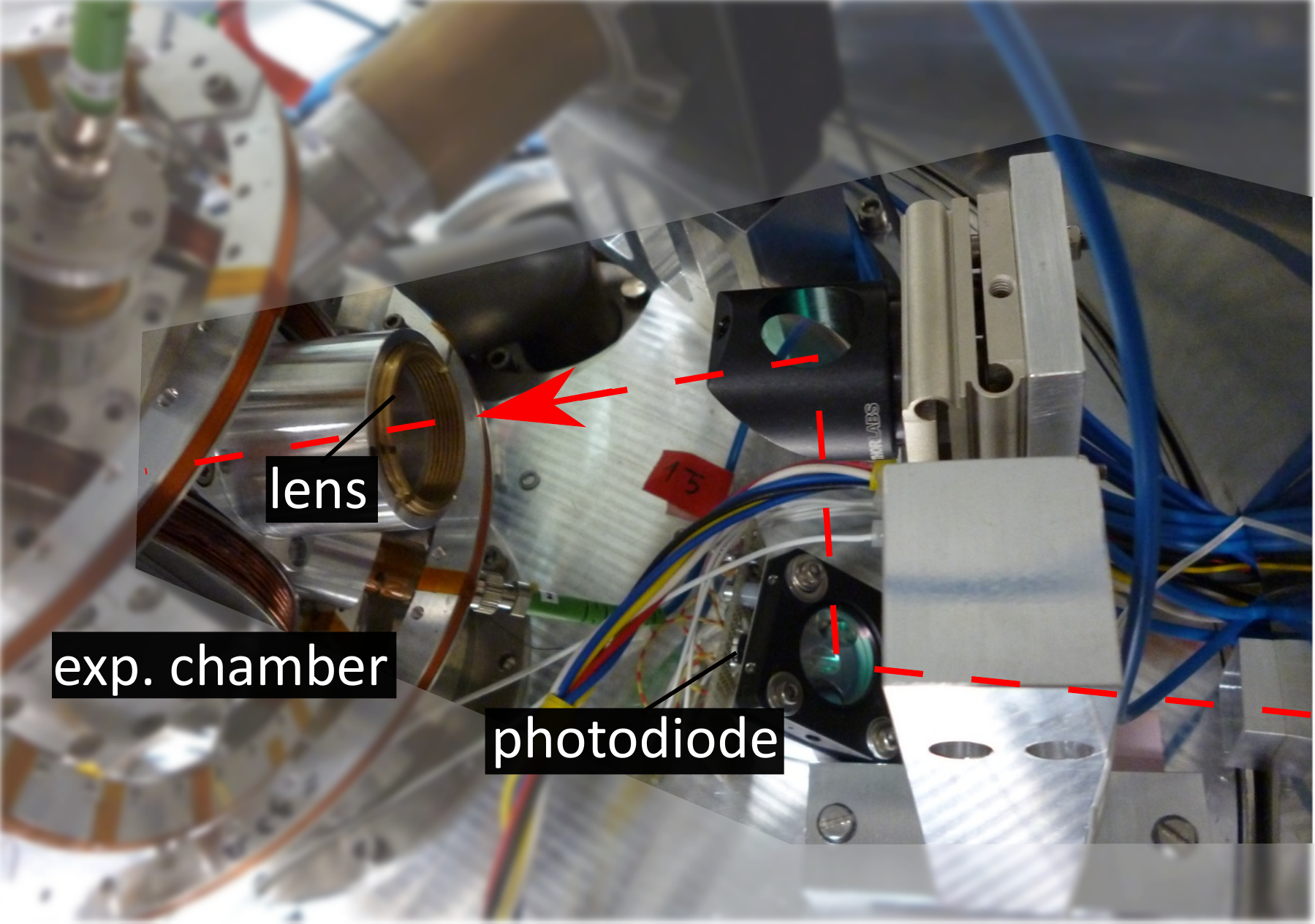}
}
\caption{Photo shows the coupling system for the dipole laser.}
\label{Couple}
\end{figure}

A free-beam setup guides the laser beam to the experimental chamber. Here, optics with a HR coating for $2\,\muup{\text{m}}$ are used to minimize the light absorption. At the vacuum chamber an adjustable mirror and a lens ($f=75$\,mm) are used to focus the laser beam into the middle of the experimental chamber (see Fig. \ref{Couple}). A photodiode at the back side of one mirror is used for the closed-loop control.

Since the electrical power requirements for the fiber laser system are challenging for drop-tower experiments, this laser system is supplied by a separate power supply. At full load the two temperature controllers (\textit{LAIRD Technologies}, TC-XX-PR-59) for oscillator and amplifier take up $120$\,W of power and together with the energy consumption of the two current drivers (oscillator: custom design, amplifier: \textit{PicoLas}, LDP-C80-20) of $213$\,W it yields $333$\,W. Two 12\,V-lead-bat\-teries connected in series with $16$\,Ah can buffer this amount for more than one hour.

\section{First test in microgravity}
\label{test}
\begin{figure}
\centering
\resizebox{0.3\textwidth}{!}{
  \includegraphics{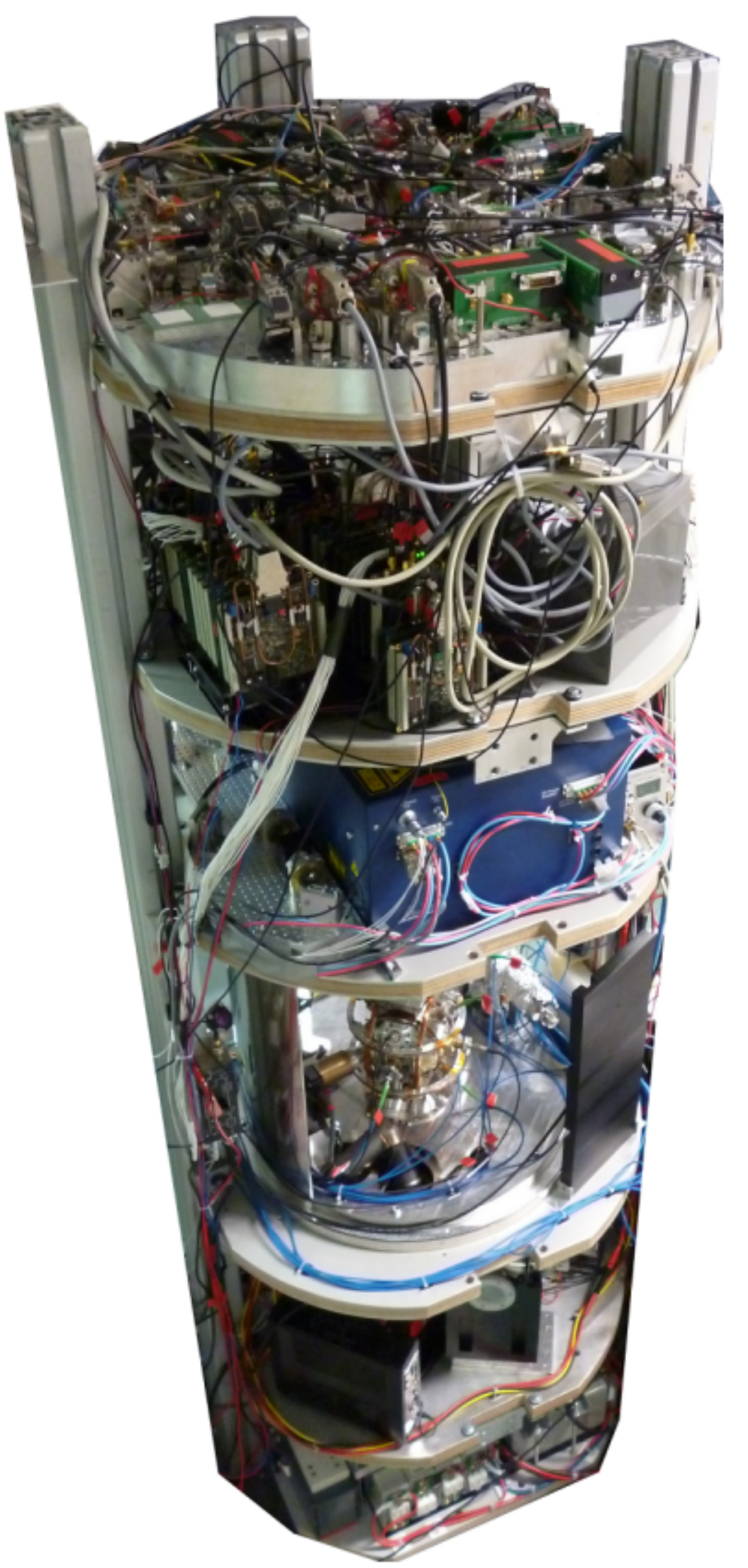}
}
\caption{This picture shows the drop capsule for performing experiments with cold atoms in microgravity.}
\label{capsule4}
\end{figure}
\begin{figure}
\centering
\resizebox{0.48\textwidth}{!}{
  \includegraphics{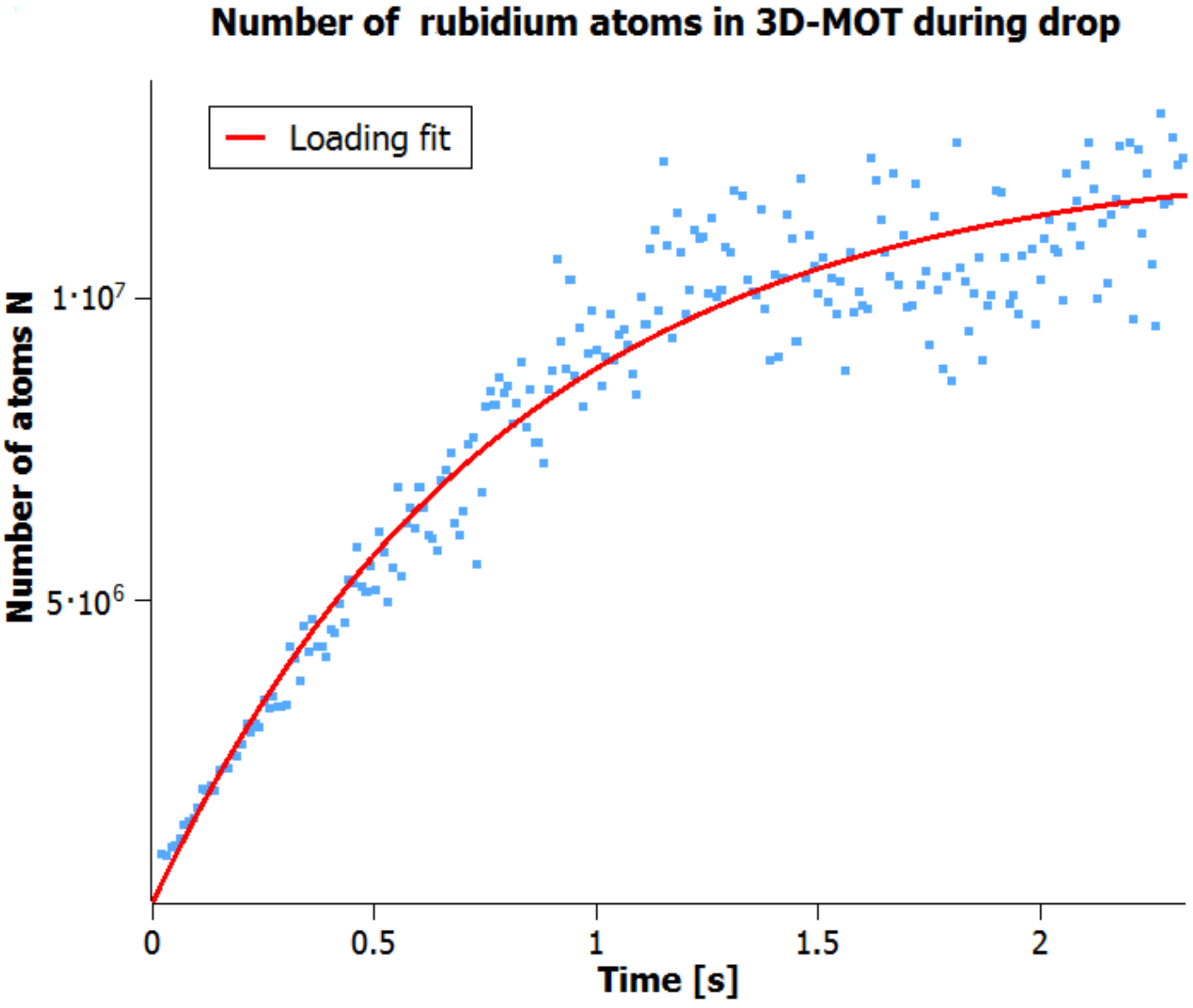}
}
\caption{Loading rate of $^{87}{\text{Rb}}$-MOT in microgravity. Shortly after release ($\approx 30$\,ms) the 3D-MOT was loaded.}
\label{num0g}
\end{figure}
\begin{figure}
\centering
\resizebox{0.48\textwidth}{!}{
  \includegraphics{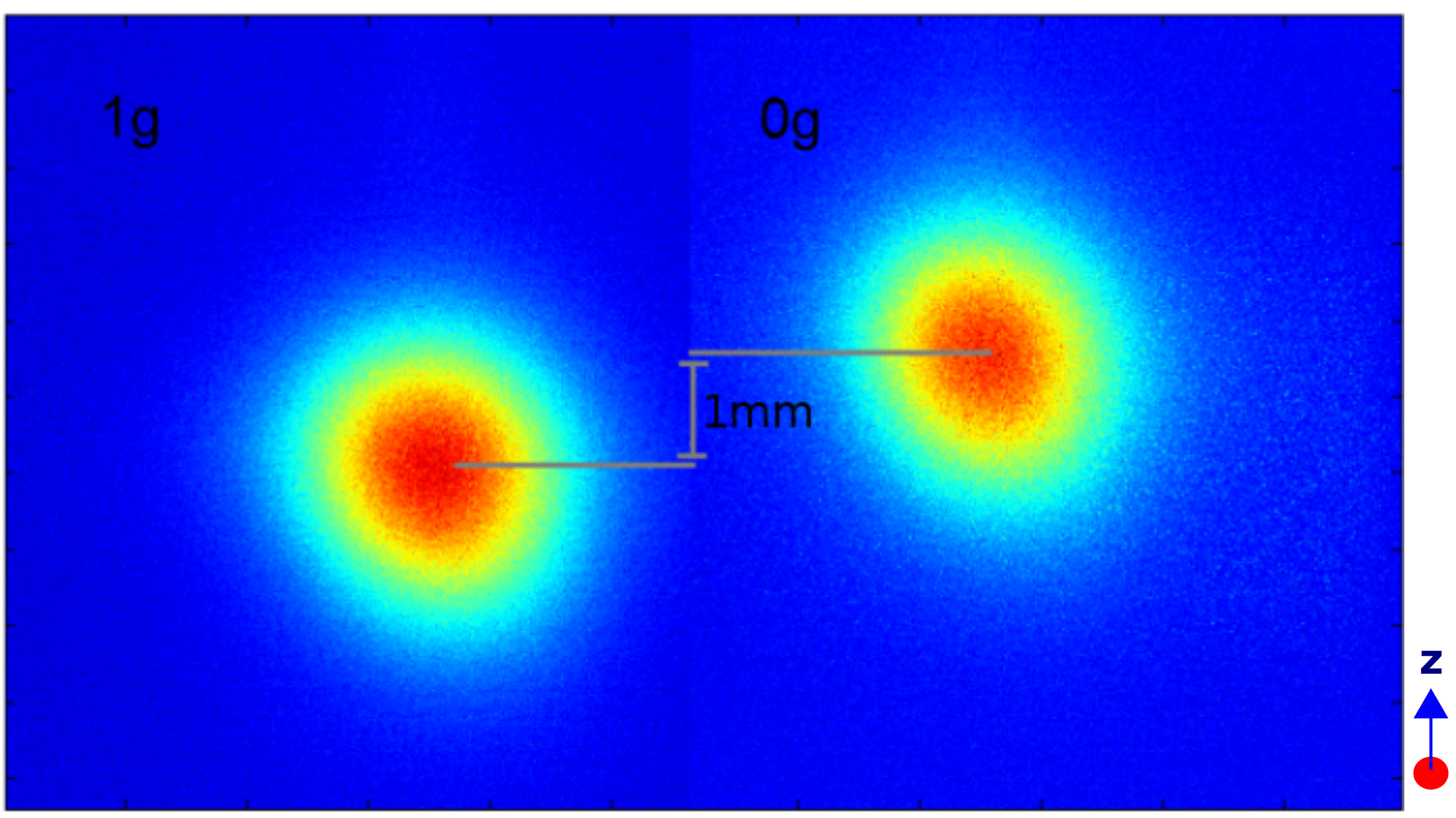}
}
\caption{The figure on the right side shows the camera picture of a rubidium ensemble in false color representation which was recorded from the MOT during microgravity. For comparison the left side shows a picture shortly before the release in the droptower. Due to gravity the atoms on the left picture are closer to the bottom of the chamber (-z direction).}
\label{Vergleich}
\end{figure}

The drop capsule (see Fig. \ref{capsule4}) including the setup, shown in Figure \ref{Setup}, was successfully tested in microgravity with respect to general hard- and software performance. 30\,ms after the release of the capsule a Rb-MOT was loaded for 2.5\,s in microgravity. This was followed by a time of flight (TOF) with $t_{\text{TOF}}=15$\,ms where the MOT was switched off by shutting down the current of the magnetic coils. Fluorescence from the atomic ensemble was detected by means of the photodiode (see Fig. \ref{num0g}) and the CCD camera as well. The atom number, determined by the photodiode signal, reached a maximum of $N_{\text{Rb0g}}=1.17\,\times\,10^{7}$. Figure \ref{Vergleich} shows the MOT fluorescence detected in microgravity as compared to a MOT created under 1\,g just before the release of the drop capsule. The resulting offset of 1\,mm in vertical position is clealy visible. The maximum atom number in the MOT under gravity had the value $N_{\text{Rb1g}}=1.7\times10^{7}$. These performances are on the same scale but differ from the performance in the the laboratory (see Fig. \ref{MOT}). The reason for this was less cooling power for the 3D-MOT (75\%) compared to the value in Table \ref{tab}. During the drop preparation the value for the cooling light power decreased due to a not locked adjuster at one mirror holder. 
\newline
The pressure as recorded by the ion getter pump current increased up to $4 \times 10^{-10}$mbar right after deceleration but settled down to a level of $7 \times 10^{-11}$mbar within 50\,s.\newline
After increasing the number of trapped potassium atoms in the laboratory described above a $\mu$g test with potassium will follow in the near future. 

\section{Conclusion and Outlook}
\label{conclusion}
We have described the technical realization of a compact system for future cold atom experiments to be used in the drop tower Bremen. The setup features a miniaturized laser system, which delivers the light for cooling $^{87}{\text{Rb}}$ and $^{39}{\text{K}}$ atoms in a combined MOT-system. The system reaches ultra-high vacuum conditions and provides $> 10^{8}$ $^{87}{\text{Rb}}$ atoms at a temperature of $145\,\muup$K, which was also demonstrated in microgravity. Nearly $10^{7}$ $^{39}{\text{K}}$ atoms can be trapped simultaneously in the laboratory. Our observations indicate that the number of trapped potassium atoms is limited by the laser power. Thus, the laser system will be modified to deliver more optical power. Also an improvement of the potassium partial pressure by increasing the oven temperature will be attempted. This however has to be traded against the increase in background pressure of the system due to the higher oven temperature.

A special feature of the setup is that it foresees the use of an optical dipole trap as opposed to an atom chip as applied in previous microgravity experiments. The implementation of such a dipole trap in our setup will allow for the advantages of evaporative cooling in a free-space trap with full optical access, at the cost of more challenging power and space requirements. Our setup meets these requirements and in particular allows to operate the necessary high-power laser within the constraints of a drop-capsule setup.

The next step will be to implement the evaporative cooling for rubidium by the use of the hybrid optical trap in the lab and in microgravity. Ultimately we aim to cool both species to sub-$\muup$K temperature by using a Feshbach resonance for sympathetic cooling, thus establishing a source for dual-species interferometry measurements in microgravity. The comparison of the free fall of $^{87}{\text{Rb}}$ and $^{39}{\text{K}}$ within the drop tower may thus enable a single-shot sensitivity of $\eta_{\text{est}}< 2.5\times 10^{-9}$. Beyond this, the limits of the drop tower clearly are in terms of statistics and study of systematics due to the limited number of measurements per day that can be taken. These may be overcome in future space-based microgravity experiments that harvest the full potential of matter-wave interferometry with extended free-fall times.

\begin{acknowledgements}
We acknowledge support by the German Space Agency DLR with funds provided by the Federal Ministry for Economic Affairs and Energy (BMWi) under grant number DLR 50 WM 1142.
\end{acknowledgements}

\end{document}